\documentclass[prc,superscriptaddress,showpacs,floatfix,widetables]{revtex4}
\usepackage[dvips]{graphicx}
\usepackage{epsfig}
\usepackage{psfig}
\usepackage{pst-plot}
\usepackage{bm}
\usepackage{mathrsfs}
\usepackage{amsfonts}
\usepackage{amssymb}
\usepackage{dcolumn}
\usepackage{lscape}
\usepackage{graphicx}

\begin{document}

\title{Effective interactions and shell model studies of heavy tin isotopes}

\author{M.~P.~Kartamyshev, T.~Engeland, M.~Hjorth-Jensen, and E.~Osnes}

\affiliation{Department of Physics and Centre of Mathematics for Applications, 
University of Oslo, N-0316 Oslo, Norway}

\date{\today}

\begin{abstract}
We present results from large-scale shell-model calculations of even and odd tin isotopes from
$^{134}$Sn to $^{142}$Sn with a
shell-model space defined by the  $1f_{7/2},2p_{3/2},0h_{9/2},2p_{1/2},1f_{5/2},0i_{13/2}$ single-particle orbits.
An effective two-body interaction based on modern nucleon-nucleon interactions is employed.
The shell-model results are in turn analyzed for their pairing content using a generalized 
seniority approach. Our results indicate that a pairing-model picture captures a great deal of the 
structure and the correlations of the lowest lying states for even and odd isotopes. 
\end{abstract}


\maketitle

\section{Introduction}
The region of neutron-rich nuclei around 
$^{132}$Sn with the number of protons at or just below the $Z=50$ shell closure, 
is an area of current 
experimental and theoretical interest, see for example
Refs.~\cite{bhattacharyya2001,radford2002,hribf,dillman2003, jacob2002,terasaki2002,stone2005,shergur2002,shergur2005a,shergur2005b}. 
Recent results span from $\beta$ and 
$\gamma$ spectroscopy at CERN/Isolde of the $N=82$ $r$-process  ``waiting point'' 
nucleus $^{130}$Cd \cite{dillman2003}, 
with several interesting astrophysical implications,  
to the measurement of the transition strengths $B(E2; 0^+\rightarrow 2^+)$ and nuclear moments of
Te and Sn isotopes at the HRIBF facility at Oak Ridge National Laboratory \cite{radford2002,hribf,stone2005}. 
For the latter experiments, 
the theoretical analysis of Ref.~\cite{terasaki2002}, using a quasiparticle random phase approximation,  
pointed to a weakened pairing strength 
for the interacting neutrons in order to explain the experimental results. 

A series of experiments at CERN/Isolde \cite{is378} have aimed at decay studies of  
$^{135 - 140}$Sn via  laser ionization techniques, in order to obtain
half-lives and $\gamma$-ray spectra for these very neutron-rich Sn nuclides
that lie directly in the path of the r-process. 
In Ref.~\cite{shergur2002} data were presented for $T_{1/2}$ and neutron emission probabilities for 
$^{135-137}$Sn.
Recently, data from various $\beta$-decays of  
$^{134,135}$Sn were used to confirm and establish new levels in $^{134}$Sb and $^{135}$Sb \cite{shergur2005a,shergur2005b}.
Data for beta-delayed neutron decays of $A=137$ and $A=138$ have also been obtained.
These and other data are of importance for our understanding of the presence of a single site for the $r$-process
nucleosynthesis, or at least a single mechanism for elements above $Z = 56$ and
possibly a second site/mechanism for elements below $Z = 56$, see for example the discussion in Ref.~\cite{dillman2003}.

The lifetime of the isotopes $^{136-138}$Sn are of the order of few milliseconds \cite{is378,shergur2002}, a fact 
which poses severe limitations to our capability of obtaining spectroscopic information with present laser ionization
techniques. Thus, 
except for the above mentioned $^{134}$Sn isotope and the low-energy spectrum of
$^{133}$Sn \cite{hoff96}, data for heavy tin isotopes ($A=134-142$) are either
rather limited or just missing. Theoretical analyses, 
except those reported in Refs.~\cite{shergur2002,shergur2005a,shergur2005b} and a 
study of two and three valence particles by Coraggio {\em et al} \cite{gigi2002}, provide
little spectroscopic information for nuclei beyond $^{135}$Sn. The sole exception we are aware of are the
mean-field investigations of Hoffman and Lenske \cite{hoffmann} and Dobaczewski {\em et al} \cite{doba1996}.  
A theoretical analysis of the spectra of these nuclei can therefore provide several possible guidelines
for interpretations of data of relevance 
for astrophysical studies of the r-process path, 
and  in particular for
investigation of spectroscopic trends when one moves towards the neutron drip line.

It is within the latter topic we focus our attention here. 
The rationale for the present study, in addition to providing spectroscopic information on 
even and odd Sn isotopes for $A=134-142$, 
is to analyze the pairing structure of the low-lying  
states of these nuclei as we move towards the drip line (the reader should note that the drip line for the Sn isotopes is predicted at
$A\sim 160$ \cite{doba1996}).
The structure of pairing correlations and the strength of the pairing interaction as one
moves towards the drip line, 
in addition to our understanding
of how shells evolve, is a topic
of great experimental and theoretical interest in low-energy nuclear physics. 

To provide the above mentioned information on pairing correlations, 
we perform large-scale shell-model calculations using 
realistic effective interactions. 
We present results from extensive shell model studies of heavy Sn isotopes with up to $10$
valence neutrons beyond the $^{132}$Sn core. The effective neutron-neutron interaction is tailored for 
a shell-model space which  includes the single-particle states
$1f_{7/2},2p_{3/2},0h_{9/2},2p_{1/2},1f_{5/2}$ and $0i_{13/2}$. A perturbative many-body scheme
is employed to derive the effective interaction starting with a realistic model for the 
free nucleon-nucleon interaction, 
as detailed in Ref.~\cite{hko95} and reviewed briefly in the next section. 
The resulting wave functions of selected low-lying states of odd and even
Sn isotopes are in turn analyzed for their pairing content using the generalized 
seniority approach developed by Talmi \cite{talmi1993}.
This provides one possible way of extracting pairing correlations and information 
about the pairing strength as one increases the number
of valence nucleons.

This work falls in four sections. In section \ref{eff_int_and_SM} we give a 
brief outline of the
theoretical framework for obtaining an effective interaction for the nuclei of interest. 
Some basic shell-model features are 
also presented.
The results from shell-model calculations for both even and odd isotopes 
are presented in section
\ref{results} together with a generalized seniority analysis. 
Concluding remarks are found in section \ref{conclusions}.

\section{Shell-model Hamiltonian }\label{eff_int_and_SM}

We give here a brief review of our calculational recipe, for further details see
Ref.~\cite{hko95}. 
Our scheme to obtain an effective two-body interaction appropriate for 
heavy tin isotopes starts with a free nucleon-nucleon  interaction $V$
appropriate for nuclear physics at low and intermediate energies. 
Here we employ the potential model of Ref.~\cite{cdbonn}, the so-called charge-dependent Bonn interaction. 
This is  an extension of the  one-boson-exchange models of the Bonn group \cite{mac89},
where mesons like $\pi$, $\rho$, $\eta$, $\delta$, $\omega$ and the fictitious
$\sigma$ meson are included.
Thereafter we need to handle 
the fact that the strong repulsive core of the nucleon-nucleon potential $V$
is unsuitable for perturbative approaches. This problem is overcome
by introducing the reaction matrix $G$ given by the solution of the
Bethe-Goldstone equation
\begin{equation}
    G=V+V\frac{Q}{\omega - H_0}G,
\end{equation}
where $\omega$ is the unperturbed energy of the interacting nucleons,
and $H_0$ is the unperturbed hamiltonian. The operator $Q$, commonly referred to
as the Pauli operator, is a projection operator which prevents the
interacting nucleons from scattering into states occupied by other nucleons.
In this work we solve the Bethe-Goldstone equation for five starting
energies $\omega$, by way of the so-called double-partitioning scheme
discussed in for example Ref~\cite{hko95}. The $G$-matrix is the sum over all
ladder type diagrams. This sum is meant to renormalize the strong repulsive
short-range part of the interaction. The physical interpretation is that the
particles must interact with each other an infinite number of times in order
to produce a finite interaction. 

Since the $G$-matrix represents
just the summation to all orders of ladder diagrams with particle-particle
diagrams, there are obviously other terms which need to be included in an effective
interaction. Long-range effects represented by core-polarization terms are also needed.
The first step then is to define the so-called $\hat{Q}$-box given by
\begin{equation}
   P\hat{Q}P=PGP + P\left(G\frac{Q}{\omega-H_{0}}G\\ + G
   \frac{Q}{\omega-H_{0}}G \frac{Q}{\omega-H_{0}}G +\dots\right)P.
   \label{eq:qbox}
\end{equation}
The $\hat{Q}$-box is made up of non-folded diagrams which are irreducible
and valence linked.
We can then obtain an effective two-body interaction
$H_{\mathrm{eff}}=\tilde{H}_0+V_{\mathrm{eff}}$ in terms of the $\hat{Q}$-box,
with \cite{hko95}
\begin{equation}
    V_{\mathrm{eff}}(n)=\hat{Q}+{\displaystyle\sum_{m=1}^{\infty}}
    \frac{1}{m!}\frac{d^m\hat{Q}}{d\omega^m}\left\{
    V_{\mathrm{eff}}(n-1)\right\}^m,
    \label{eq:fd}
\end{equation}
where $(n)$ and $(n-1)$ refer to the effective interaction after $n$ and $n-1$ 
iterations. The zeroth iteration is represented by just the $\hat{Q}$-box. 
Observe also that the effective interaction $V_{\mathrm{eff}}(n)$ is 
evaluated at a given model space energy $\omega$, as is the case for the 
$G$-matrix as well. Here we choose $\omega =-20$ MeV.
Less than $10$ iterations were needed 
in order to obtain a numerically stable result. All non-folded diagrams through 
third order in the interaction $G$ are included. 
A harmonic oscillator basis was used to derive the single-particle radial wave functions
with an oscillator energy $\hbar\omega = 7.87$ MeV. 
For further details, see 
Ref.\ \cite{hko95}.  

The present neutron-neutron interaction, the corresponding proton-proton and the proton-neutron interactions
derived for holes and particles states using $^{132}$Sn as closed shell core, have 
been employed in several shell-model calculations of nuclei around $^{132}$Sn, see for example
Refs.~\cite{dillman2003,shergur2005a,shergur2005b,brown2005,stone1998,isotones1997}.

The effective two-particle interaction can in turn be used in shell-model
calculations. The shell-model problem requires the solution of a real symmetric
$n \times n$ matrix eigenvalue equation
\begin{equation}
       \tilde{H}\left | \Psi_k\right\rangle  =  E_k \left | \Psi_k\right\rangle ,
       \label{eq:shell_model}
\end{equation}
with $k = 1,\ldots, K$. At present our basic approach to finding solutions to 
Eq.\ (\ref{eq:shell_model}) is the Lanczo's algorithm; an iterative method which 
gives the solution of the lowest eigenstates. This method was 
already applied to nuclear physics problems by Whitehead {\sl et al.} 
in 1977. The technique is described in detail in Ref.\ \cite{whit77} and implemented via the Oslo shell-model code
\cite{torgeir1991}. 
The eigenstates of Eq.~(\ref{eq:shell_model}) are
written as linear combinations of Slater determinants in the $m$-scheme,
distributing the $N$ valence neutrons in all possible
ways through the single particle $m$-scheme orbitals of the 
model space,  
$1f_{7/2},2p_{3/2},0h_{9/2},2p_{1/2},1f_{5/2}$ and $0i_{13/2}$.
As seen in Table \ref{tab:table1}, the dimensionality
$n$ of the eigenvalue matrix $\tilde{H}$ is increasing rapidly
with increasing number of valence particles, and
for the Sn isotopes of interest can take values up to $n \approx 6 \times 10^{7}$.
\begin{table}[htbp]
\begin{center}
\caption{Total number of basis states for the shell-model calculation 
with the single-particle orbitals
$1f_{7/2},2p_{3/2},0h_{9/2},2p_{1/2},1f_{5/2}$ and $0i_{13/2}$
defining the model space.}
\begin{tabular}{lrcrlr}
\\
\hline
System & Dimension      & System & Dimension     & System & Dimension \\
\hline
$^{134}$Sn  & 62    &    $^{137}$Sn &     34 804 &    $^{140}$Sn & 4 606 839  \\
$^{135}$Sn  & 448   &    $^{138}$Sn &   207 514  &    $^{141}$Sn & 17 574 855 \\ 
$^{136}$Sn  & 4 985 &    $^{139}$Sn & 1 049 533  &    $^{142}$Sn & 59 309 576 \\ 
\hline
\end{tabular}
\label{tab:table1}
\end{center}
\end{table}

\begin{figure}[htbp]
\setlength{\unitlength}{1.4cm}
\begin{center}
\begin{picture}(2,4)(0,-1)
\newcommand{\lc}[1]{\put(0,#1){\line(1,0){1}}}
\newcommand{\ls}[2]{\put(1.9,#1){\makebox(0,0){{\scriptsize $#2$}}}}
\newcommand{\lsr}[2]{\put(2,#1){\makebox(0,0){{\scriptsize $#2$}}}}
\put(-.2,3.4){\makebox(0,0){\large MeV}}
\thicklines
\put(-.75,-.5){\line(0,1){4}}
\multiput(-.75,.0)(0,1){4}{\line(1,0){.1}}
\multiput(-.75,.5)(0,1){3}{\line(1,0){.05}}
\put(-0.55,2){\makebox(0,0){2}}
\put(-0.55,1){\makebox(0,0){1}}
\put(-0.55,0){\makebox(0,0){0}}
\lc{0.0000}    \ls{0.0000}{7/2- \;\;(0.0000)}
\lc{0.8537}   \ls{0.8537}{3/2-  \;\;(0.8537)}
\lc{1.5609}   \ls{1.4500}{9/2-  \;\;(1.5609)}
\lc{1.6557}   \ls{1.6800}{1/2-  \;\;(1.6557)}
\lc{2.0046}   \ls{2.0000}{5/2-  \;\;(2.0046)}
\lc{2.6940}   \ls{2.6940}{13/2+ \;\;(2.6940)}
\put(0.5,-.3){\makebox(0,0){{\large $^{133}$Sn}}}
\end{picture}
\end{center}
\caption{Adopted single-particle energies for the orbitals
$1f_{7/2},2p_{3/2},0h_{9/2},2p_{1/2},1f_{5/2}$ and $0i_{13/2}$
in shell model calculations.}
\label{fig:fig1}
\end{figure}
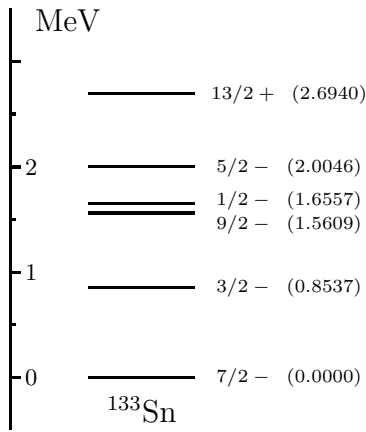

The single-particle energies are extracted from the experimental 
$^{133}$Sn spectrum \cite{hoff96}, except for the $0i_{13/2}$ single-particle
orbital which lies above the $^{132}$Sn $+\,\,n$ threshold 2.45(5) MeV \cite{mezilev}. 
If one interprets the 2434-keV level in $^{134}$Sb as the $(\pi g_{7/2}, \nu i_{13/2}) 10^+$
configuration \cite{urban99},  the position of the $0i_{13/2}$ orbital can be estimated to 
$2.6940 \pm 0.2$ MeV. The adopted single-particle energies used in this work are displayed
in Fig.\ \ref{fig:fig1}.\newline\newline\newline

We anticipate parts of the discussion of our results by pointing to the fact that 
there is a difference  of 0.85 MeV between 
the ground state in $^{133}$Sn and the first excited state $(3/2)^-$ and 1.56 MeV to the
next excited state with spin and parity assignnement  $(9/2)^-$. Furthermore, the shell gap for neutrons
is
\begin{eqnarray}
\Delta \epsilon_{\nu} & = & \epsilon_{\nu}(1f_{7/2}) - \epsilon_{\nu}(0g_{7/2})
\nonumber \\
& = &
[\mathrm{BE}(^{132}\mathrm{Sn})-\mathrm{BE}(^{133}\mathrm{Sn})]
+ [\mathrm{BE}(^{132}\mathrm{Sn})-\mathrm{BE}(^{131}\mathrm{Sn})]
\nonumber \\
& = & 4.84 \hspace{0.1cm}\mathrm{MeV}.
\label{eq:shellgap}
\end{eqnarray}
The first excited negative parity state in $^{132}$Sn has an excitation energy of 
$4.351$ MeV and total angular momentum $J^{\pi}=3^-$. This indicates that the contribution from the two-body interaction
is of the order of $\sim 0.5$ MeV if we assume a particle-hole model space consisting of the above mentioned
single particle states. Compared with closed-shell cores like $^{16}$O, $^{40}$Ca and $^{56}$Ni, the cross-shell
interaction for the present study is much weaker, almost an order of magnitude compared with a similar analysis
for $^{16}$O (in $^{16}$O the shell gap for neutrons is 11.521 MeV and the first excited negative parity state
$J^{\pi}=3^-_1$ is located at 6.129 MeV). This fact applies to the matrix elements of the derived
two-body interaction as well, an expected feature since the valence neutrons are on average further apart from
each other. An investigation of such trends of the effective interactions 
for nuclei from $^{4}$He to $^{208}$Pb was done
in Ref.~\cite{jpg22}. There the authors demonstrated that the interaction matrix elements for valence particle
systems become on average smaller with increasing mass number $A$.

In the analysis of the pairing content of our results, we will see that the interplay 
between the single-particle spacing in $^{133}$Sn and the strength of the two-body interaction
allows a qualitative understanding of the reported results.

\section{Results and discussion}\label{results}

In this section we present results of the shell-model calculations 
for Sn isotopes with up to $10$ valence neutrons
beyond the $^{132}$Sn closed-shell core. As the experimental data 
available at the present time for the $A \ge 134$
tin isotopes are quite sparse, we have a limited possibility to check the validity of our effective 
interaction in this region of the nuclear chart. Hence, most of the present results are of a predictive
character, providing shell-model guidelines for future experiments of the $A \ge 134$ tin isotopes.

\subsection{Even tin isotopes}

The results of the shell-model calculation for the even tin isotopes 
are displayed in Table \ref{tab:even_isotopes}
for selected states. We obtain a rather
good agreement with the measured energies of the positive parity 
yrast states in $^{134}$Sn \cite{shergur2002,korgul2000}. 
For $^{142}$Sn we have calculated only the positive parity states. For this nucleus the number
of basis states becomes very large (see Table \ref{tab:table1}), 
mainly due to the contributions from the $0i_{13/2}$ 
negative parity orbital. 
As a consequence, the shell model calculations for $^{142}$Sn are rather time consuming. 
However, if one puts limits on the number of particles which can occupy a particular 
orbital, 
the number of basis states can be substantially reduced. Thus, for this nucleus we choose 
the upper limit of particle number for 
the $0h_{9/2}, 1f_{5/2}$ and $0i_{13/2}$ orbitals to be respectively ly 4, 2 and 2. This reduction allows
the calculations to be performed in a reasonable time, while introducing only a 
negligible error to the energies of the 
nuclear states of $^{142}$Sn.

\begin{table}[htbp]
\begin{center}
\caption{Selected low-lying states for  $^{134}$Sn, $^{136}$Sn, $^{138}$Sn, $^{140}$Sn and $^{142}$Sn.
Experimental data for $^{134}$Sn are taken from \cite{shergur2002,korgul2000}. All entries in MeV.}
\begin{tabular}{cccc cccc cccc cccc cccc cccc}
\hline
&&&&&&&&&&&&&&&&&&&&\\
\multicolumn{4}{c}{ $^{134}$Sn} &
\multicolumn{4}{c}{ $^{136}$Sn} &
\multicolumn{4}{c}{ $^{138}$Sn} &
\multicolumn{4}{c}{ $^{140}$Sn} &
\multicolumn{4}{c}{ $^{142}$Sn} &\\
&{$J^{\pi}_i$}&Theory &Experiment &&
&{$J^{\pi}_i$}&Theory &&
&{$J^{\pi}_i$}&Theory &&
&{$J^{\pi}_i$}&Theory &&
&{$J^{\pi}_i$}&Theory &\\
\hline
& $0^{+}_{1}$ & 0.0000   &0.0000 &&   &$0^{+}_{1}$ & 0.0000  &&   &$0^{+}_{1}$ & 0.0000 &&   &$0^{+}_{1}$ & 0.0000 && &$0^{+}_{1}$& 0.0000  &\\
& $0^{+}_{2}$ & 2.2822   & &&   &$0^{+}_{2}$ & 1.8779  &&   &$0^{+}_{2}$ & 1.5280 &&   &$0^{+}_{2}$ & 1.2256 && &$0^{+}_{2}$& 1.0568  &\\
& $1^{+}_{1}$ & 2.4141   & &&   &$1^{+}_{1}$ & 1.9571  &&   &$1^{+}_{1}$ & 2.0327 &&   &$1^{+}_{1}$ & 2.0539 && &$1^{+}_{1}$& 1.6878  &\\
& $2^{+}_{1}$ & 0.7748&0.7256&& &$2^{+}_{1}$ & 0.7339  &&   &$2^{+}_{1}$ & 0.7615 &&   &$2^{+}_{1}$ & 0.8028 && &$2^{+}_{1}$& 0.7457  &\\
& $2^{+}_{2}$ & 1.6601   & &&   &$2^{+}_{2}$ & 1.4642  &&   &$2^{+}_{2}$ & 1.2840 &&   &$2^{+}_{2}$ & 1.3293 && &$2^{+}_{2}$& 1.2221  &\\
& $3^{+}_{1}$ & 2.0978   & &&   &$3^{+}_{1}$ & 1.8183  &&   &$3^{+}_{1}$ & 1.6072 &&   &$3^{+}_{1}$ & 1.5394 && &$3^{+}_{1}$& 1.5687  &\\
& $4^{+}_{1}$ & 1.1161&1.0734&& &$4^{+}_{1}$ & 1.1614  &&   &$4^{+}_{1}$ & 1.3550 &&   &$4^{+}_{1}$ & 1.4381 && &$4^{+}_{1}$& 1.5057  &\\
& $4^{+}_{2}$ & 1.9486   & &&   &$4^{+}_{2}$ & 1.3332  &&   &$4^{+}_{2}$ & 1.5171 &&   &$4^{+}_{2}$ & 1.6308 && &$4^{+}_{2}$& 1.5620  &\\
& $5^{+}_{1}$ & 2.1708   & &&   &$5^{+}_{1}$ & 1.6877  &&   &$5^{+}_{1}$ & 1.8158 &&   &$5^{+}_{1}$ & 1.7667 && &$5^{+}_{1}$& 1.7471  &\\
& $6^{+}_{1}$ & 1.2582&1.2474&& &$6^{+}_{1}$ & 1.3770  &&   &$6^{+}_{1}$ & 1.5288 &&   &$6^{+}_{1}$ & 1.8969 && &$6^{+}_{1}$& 1.8155  &\\
& $6^{+}_{2}$ & 2.6215   & &&   &$6^{+}_{2}$ & 2.1620  &&   &$6^{+}_{2}$ & 2.1359 &&   &$6^{+}_{2}$ & 2.0722 && &$6^{+}_{2}$ & 2.0019&\\
& $7^{+}_{1}$ & 2.9538   & &&   &$7^{+}_{1}$ & 2.2915  &&   &$7^{+}_{1}$ & 2.1221 &&   &$7^{+}_{1}$ & 2.1126 && &$7^{+}_{1}$ & 2.0791\\
& $8^{+}_{1}$ & 2.4634&2.5089&& &$8^{+}_{1}$ & 2.1189  &&   &$8^{+}_{1}$ & 2.3606 &&   &$8^{+}_{1}$ & 2.4042 && &$8^{+}_{1}$ & 2.3835\\
& $8^{+}_{2}$ & 4.6571   & &&   &$8^{+}_{2}$ & 2.4125  &&   &$8^{+}_{2}$ & 2.7503 &&   &$8^{+}_{2}$ & 2.6369 && &$8^{+}_{2}$ & 2.7592\\
\hline
\end{tabular}
\label{tab:even_isotopes}
\end{center}
\end{table}
We have omitted negative parity states. In our case they are constructed from particle excitations and the first states
appear at $\sim 4-5$ MeV. This is also a region where we expect, from the shell gap of Eq.~(\ref{eq:shellgap}), negative parity 
particle-hole excitations to appear.
  
Of interest here is the fact the $0_1^+-2_1^+$ spacing remains nearly constant, except for a small increase at
$^{140}$Sn due to the filling of the $1f_{7/2}$ single-particle orbit. This constancy is similar to what we have for
the lighter even tin isotopes, from $^{102}$Sn to $^{130}$Sn. The difference is that the gap is smaller by approximately
$0.5$ MeV. Compared with the matrix elements for the model space of the lighter tin isotopes 
(the model space for   $^{102}$Sn to $^{130}$Sn consists of the single-particle orbits
$0g_{7/2},2s_{1/2},0h_{11/2},1d_{3/2},1d_{5/2}$), the 
interaction matrix elements are on average smaller for the present  model space. This is simply due to the fact 
that the nucleons are farther apart on average from each other than they are 
for the lighter tin isotopes. 
For the $0_1^+-2_1^+$ spacing, 
the most important configurations are those where the single-particle orbits $2p_{3/2}$ and $1f_{7/2}$ are involved while the
$0h_{9/2}$ plays a  role in excited states with higher spin values, 
as expected from the 
single-particle spectrum of Fig.~\ref{fig:fig1}. 
Actually, the gross features of both the even and odd low-energy spectrum can be captured by an interaction
defined  for a model space including only the  $2p_{3/2}$ and $1f_{7/2}$ orbits. This feature is similar to what we have in 
the calcium isotopes, except that the $1p_{3/2}$ and $0f_{7/2}$ orbits have one node less. These properties
of the calcium isotopes were discussed more than forty years ago, see for example Refs.~\cite{eivind1965,eivind1966}.

The matrix elements which involve the $2p_{3/2}$ and $1f_{7/2}$ orbits oscillate in absolute value 
between $\sim 0.5-1.0$ MeV. This means, if we adopt a simple pairing model analysis, as done in Ref.~\cite{dhj2003}, that the relation
between the single-particle spacing and the interaction matrix elements is of the order of $\sim 0.5-1.0$. 
Within the framework of a simple pairing model, this is a region where we expect pairing correlations to play a dominant role. In this case, 
the interaction determines much of the pairing gap, viz the difference between the seniority zero ground state and the first excited seniority
two state, where we have one broken pair.
As we will show below, a generalized seniority analysis indicates indeed that pairing correlations are still
strong resulting in a nearby  constant o$0_1^+-2_1^+$ spacing, as is the case for the lighter Sn isotopes. See also the discussion
in Ref.~\cite{dhj2003}. 

 The results for the $B(E2)$ calculations are shown in table \ref{tab:E2_even_isotopes}.
The values of calculated $B(E2)$ transition probabilities are presented in Weisskopf units (W.u.), defined as \cite{bm69}
\begin{equation}
  B_w (E \lambda ) = \frac{(1.2)^4}{4\pi} \left ( \frac{3}{\lambda + 3} \right )^2\!\! A^{\frac{2\lambda}{3}}\ e^2 fm^4,
\end{equation}
with $\lambda =2$ in our case.
Experimental values of $B(E2)$ for the nuclei of interest are currently available
only for two transitions in $^{134}$Sn: $B(E2: 6_1^+ \rightarrow 4_1^+)= 36 \pm 7$ e$^2$fm$^4$ or $0.88 \pm 0.17$ 
W.u.~\cite{zhang1997}  and $B(E2:0_1^+ \rightarrow 2_1^+) \approx  295.0$ e$^2$fm$^4$ or $7.24$ W.u.~\cite{radford2002}.
These transitions probabilities can be reproduced in our calculations by adjusting the effective neutron charge
value to $e_{n}^{eff} = 0.66\ e$ for  the $6_1^+ \rightarrow 4_1^+$ transition and $e_{n}^{eff} = 0.62\ e$ for
the $0_1^+ \rightarrow 2_1^+$ transition, respectively. 

One could derive state dependent effective charges along the same lines as for the effective interaction.
In this work we limit however ourselves to a constant effective charge. An average of the neutron effective charge
obtained above, $e_{n}^{eff} = 0.64\ e$, has been used for every calculated transition. 
It should be noted that the authors of 
Ref.~\cite{shergur2002} use a similar value, $e_{n}^{eff} = 0.70\ e$

The $B(E2: 2_1^+ \rightarrow 0_1^+ )$ value scale nicely with the number of valence neutrons, except for $^{142}$Sn, where 
we see a reduction due to the weak shell closure at $^{140}$Sn, see the discussion below on odd isotopes as well. 
\begin{table}[htbp]
\begin{center}
\caption{Selected $B(E2)$ values for $^{134}$Sn, $^{136}$Sn, $^{138}$Sn, $^{140}$Sn and $^{142}$Sn. 
$e_{n}^{eff}=0.64e$. All entries are in W.u.}
\begin{tabular}{cc cc cc cc cc cc}
\hline
\multicolumn{2}{c}{Transition}  & 
\multicolumn{2}{c}{ $^{134}$Sn} & 
\multicolumn{2}{c}{ $^{136}$Sn} & 
\multicolumn{2}{c}{ $^{138}$Sn} & 
\multicolumn{2}{c}{ $^{140}$Sn} &
\multicolumn{2}{c}{ $^{142}$Sn} \\
\hline
&$2_1^+ \rightarrow 0_1^+$ && 1.50   &&  2.94   && 4.19  &&   5.72 &&  5.55   \\
&$4_1^+ \rightarrow 2_1^+$ && 1.53   &&  2.04   && 0.006 &&   1.71 &&  0.78   \\
&$6_1^+ \rightarrow 4_1^+$ && 0.83   &&  0.33   && 0.27  &&   0.54 &&  0.06   \\
&$8_1^+ \rightarrow 6_1^+$ && 0.12   &&  1.25   && 0.03  &&   0.08 &&  0.04       \\
\hline
\end{tabular}
\label{tab:E2_even_isotopes}
\end{center}
\end{table}
We notice also that the $B(E2: 4_1^+ \rightarrow 2_1^+ )$ value almost vanishes at 
$^{138}$Sn. From the seniority discussion of 
subsection \ref{subsec:gensensec}, we would expect this value to be large since the seniority $\nu=2$ overlaps
with the shell-model wave functions are large for the $2^+_1$ and $4^+_1$. Within the framework of a naive 
seniority picture the corresponding transition between two seniority $\nu = 2$ states 
is expected to be larger than the value listed
in Table \ref{tab:E2_even_isotopes}. There are however other contributions as well which cancel out the naive
seniority picture.
A further  analysis and interpretation of the structure of the wave functions is 
made in connection with our generalized seniority analysis in subsection \ref{subsec:gensensec}.

\subsection{Odd isotopes}

\begin{figure}[htbp]
\setlength{\unitlength}{3.0cm}
\begin{center}
\begin{picture}(4.9,6.5)(0,-1)
\psset{xunit=1.0cm, yunit=5.5cm}
\newcommand{\drawlevel}[5]{\psline[origin={#1,#2}, linewidth=0.3pt](0,0)(1.5,0.0)\rput(#3,#4){\scriptsize \makebox(0,0){$#5$}}}
\newcommand{\connect}[4]{\psline[linewidth=0.3pt, linestyle=dotted, dotsep=1.2pt](#1,#2)(#3,#4)}
\rput(0,0){\psline(-2.0,-.3)(-2.0,2.5)}
\rput(-1.2,2.5){\makebox(0,0){\large MeV}}
\multirput(-1.85,.0)(0,1){3}{\line(1,0){.1}}
\multirput(-1.9,.5)(0,1){2}{\line(1,0){.05}}
\rput(-1.4,2){\makebox(0,0){2}}
\rput(-1.4,1){\makebox(0,0){1}}
\rput(-1.4,0){\makebox(0,0){0}}
\rput(0.1,-0.2){\makebox(0,0){{\large $^{135}$Sn}}}
\drawlevel{0.7}{0.0000}{2.0}{0.0000}{\ 7/2_1^- \ ({\bf 0.0000})}
\drawlevel{0.7}{-0.3020}{2.0}{0.3000}{\ 5/2_1^- \ (0.3020)}
\drawlevel{0.7}{-0.4168}{2.0}{0.4168}{\ 3/2_1^- \ ({\bf 0.4168})}
\drawlevel{0.7}{-0.6445}{2.0}{0.6445}{\ 3/2_2^- \ (0.6445)}
\drawlevel{0.7}{-0.7415}{2.0}{0.7415}{ 11/2_1^- \ (0.7415)}
\drawlevel{0.7}{-0.8677}{2.0}{0.8677}{\ 9/2_1^- \ ({\bf 0.8677})}
\drawlevel{0.7}{-1.0969}{2.0}{1.0900}{ 15/2_1^- \ (1.0969)}
\drawlevel{0.7}{-1.1641}{2.0}{1.1600}{\ 9/2_2^- \ (1.1641)} \connect{0.8}{1.1641}{1.13}{1.16}
\drawlevel{0.7}{-1.1884}{2.0}{1.2300}{\ 9/2_3^- \ (1.1884)} \connect{0.8}{1.1884}{1.13}{1.23}
\drawlevel{0.7}{-1.2143}{2.0}{1.3000}{\ 1/2_1^- \ ({\bf 1.2143})} \connect{0.8}{1.2143}{1.15}{1.30}
\drawlevel{0.7}{-1.7115}{2.0}{1.7115}{ 13/2_1^- \ (1.7115)}
\drawlevel{0.7}{-2.1371}{2.0}{2.1371}{ 17/2_1^- \ (2.1371)}
\rput(4.3,-0.2){\makebox(0,0){{\large $^{137}$Sn}}}
\drawlevel{-3.5}{0.0000}{6.2}{0.0000}{\ 7/2_1^- \ ({\bf 0.0000})}
\drawlevel{-3.5}{-0.2964}{6.2}{0.2964}{\ 5/2_1^- \ (0.2964)}
\drawlevel{-3.5}{-0.3707}{6.2}{0.3707}{\ 3/2_1^- \ ({\bf 0.3707})}
\drawlevel{-3.5}{-0.6069}{6.2}{0.6069}{\ 3/2_2^- \ (0.6069)} 
\drawlevel{-3.5}{-0.7250}{6.2}{0.6650}{\ 1/2_1^- \ ({\bf 0.7250})} \connect{5.0}{0.7250}{5.3}{0.6650}
\drawlevel{-3.5}{-0.7608}{6.2}{0.7200}{ 11/2_1^- \ (0.7608)} \connect{5.0}{0.7608}{5.2}{0.7300}
\drawlevel{-3.5}{-0.7754}{6.2}{0.7754}{\ 9/2_1^- \ ({\bf 0.7754})} \connect{5.0}{0.7754}{5.4}{0.7800}
\drawlevel{-3.5}{-0.8201}{6.2}{0.8300}{\ 5/2_2^- \ (0.8201)}
\drawlevel{-3.5}{-0.8860}{6.2}{0.8860}{\ 9/2_2^- \ (0.8860)}
\drawlevel{-3.5}{-0.9566}{6.2}{0.9566}{\ 7/2_2^- \ (0.9566)}
\drawlevel{-3.5}{-1.0410}{6.2}{1.0100}{\ 5/2_3^- \ (1.0410)} \connect{5.0}{1.0410}{5.4}{1.0100}
\drawlevel{-3.5}{-1.0600}{6.2}{1.0600}{\ 1/2_2^- \ (1.0600)} \connect{5.0}{1.0600}{5.4}{1.0600}
\drawlevel{-3.5}{-1.1030}{6.2}{1.1100}{\ 9/2_3^- \ (1.1030)} \connect{5.0}{1.1030}{5.4}{1.1100}
\drawlevel{-3.5}{-1.1100}{6.2}{1.1600}{\ 3/2_3^- \ (1.1050)} \connect{5.0}{1.1100}{5.4}{1.1700}
\drawlevel{-3.5}{-1.1162}{6.2}{1.2100}{\ 7/2_3^- \ (1.1062)} \connect{5.0}{1.1162}{5.3}{1.2100}
\drawlevel{-3.5}{-1.2692}{6.2}{1.2692}{ 15/2_1^- \ (1.2692)}
\drawlevel{-3.5}{-1.3884}{6.2}{1.3884}{\ 13/2_1^- \ (1.3884)}
\drawlevel{-3.5}{-1.7531}{6.2}{1.7531}{\ 17/2_1^- \ (1.7531)}
\rput(8.5,-0.2){\makebox(0,0){{\large $^{139}$Sn}}}
\drawlevel{-7.7}{0.0000}{10.4}{0.0000}{\ 7/2_1^- \ ({\bf 0.0000})}
\drawlevel{-7.7}{-0.0853}{10.4}{0.0853}{\ 3/2_1^- \ ({\bf 0.0853})}
\drawlevel{-7.7}{-0.4427}{10.4}{0.4427}{\ 1/2_1^- \ ({\bf 0.4427})}
\drawlevel{-7.7}{-0.5178}{10.4}{0.5000}{\ 5/2_1^- \ (0.5178)} \connect{9.2}{0.5178}{9.5}{0.505} 
\drawlevel{-7.7}{-0.5482}{10.4}{0.5600}{\ 3/2_2^- \ (0.5482)} \connect{9.2}{0.5482}{9.5}{0.560} 
\drawlevel{-7.7}{-0.6277}{10.4}{0.6277}{\ 9/2_1^- \ ({\bf 0.6277})}
\drawlevel{-7.7}{-0.6956}{10.4}{0.6956}{\ 7/2_2^- \ (0.6956)}
\drawlevel{-7.7}{-0.7416}{10.4}{0.7500}{\ 5/2_2^- \ (0.7416)}
\drawlevel{-7.7}{-0.8128}{10.4}{0.8128}{\ 1/2_2^- \ (0.8128)}
\drawlevel{-7.7}{-0.8548}{10.4}{0.8700}{\ 9/2_2^- \ (0.8548)} \connect{9.2}{0.8548}{9.5}{0.8700}
\drawlevel{-7.7}{-0.8633}{10.4}{0.9200}{\ 7/2_3^- \ (0.8633)} \connect{9.2}{0.8633}{9.5}{0.9200}
\drawlevel{-7.7}{-0.9185}{10.4}{0.9800}{\ 5/2_3^- \ (0.9185)} \connect{9.2}{0.9185}{9.5}{0.9800}
\drawlevel{-7.7}{-0.9766}{10.4}{1.0400}{ 11/2_1^- \ (0.9766)} \connect{9.2}{0.9766}{9.45}{1.0400}
\drawlevel{-7.7}{-1.1983}{10.4}{1.1983}{\ 13/2_1^- \ (1.1983)}
\drawlevel{-7.7}{-1.0859}{10.4}{1.0859}{\ 17/2_1^- \ (1.0859)}
\drawlevel{-7.7}{-1.6401}{10.4}{1.6401}{\ 15/2_1^- \ (1.6401)}
\rput(12.7,-0.2){\makebox(0,0){{\large $^{141}$Sn}}}
\drawlevel{-11.9}{0.0000}{14.6}{0.0000}{\ 3/2_1^- \ ({\bf 0.0000})}
\drawlevel{-11.9}{-0.1590}{14.6}{0.1590}{\ 7/2_1^- \ ({\bf 0.1590})}
\drawlevel{-11.9}{-0.4155}{14.6}{0.4155}{\ 1/2_1^- \ ({\bf 0.4155})}
\drawlevel{-11.9}{-0.5567}{14.6}{0.500}{\ 3/2_2^- \ (0.5567)}  \connect{13.4}{0.5567}{13.70}{0.500}
\drawlevel{-11.9}{-0.5645}{14.6}{0.5645}{\ 9/2_1^- \ ({\bf 0.5645})} \connect{13.4}{0.5645}{13.72}{0.5800}
\drawlevel{-11.9}{-0.6223}{14.6}{0.6223}{\ 5/2_1^- \ (0.6223)}
\drawlevel{-11.9}{-0.7035}{14.6}{0.7035}{\ 5/2_2^- \ (0.7035)}
\drawlevel{-11.9}{-0.7599}{14.6}{0.7599}{\ 7/2_2^- \ (0.7599)}
\drawlevel{-11.9}{-0.87}{14.6}{0.89}{\ 1/2_2^- \ (0.8696)} \connect{13.4}{0.8696}{13.72}{0.89}
\drawlevel{-11.9}{-0.8558}{14.6}{0.83}{\ 5/2_3^- \ (0.8558)}\connect{13.4}{0.8558}{13.72}{0.83}
\drawlevel{-11.9}{-0.9052}{14.6}{0.95}{\ 9/2_2^- \ (0.9052)}\connect{13.4}{0.91}{13.72}{0.95}
\drawlevel{-11.9}{-1.0045}{14.6}{1.0045}{\ 3/2_3^- \ (1.0045)}
\drawlevel{-11.9}{-1.2791}{14.6}{1.2791}{\ 9/2_3^- \ (1.2791)}
\drawlevel{-11.9}{-1.1046}{14.6}{1.15}{\ 11/2_1^- \ (1.1046)}\connect{13.4}{1.1}{13.72}{1.1500}
\drawlevel{-11.9}{-1.3614}{14.6}{1.3614}{\ 13/2_1^- \ (1.3614)}
\drawlevel{-11.9}{-1.7345}{14.6}{1.7345}{\ 15/2_1^- \ (1.7345)}
\drawlevel{-11.9}{-1.9114}{14.6}{1.9114}{\ 17/2_1^- \ (1.9114)}
\drawlevel{-11.9}{-1.1002}{14.6}{1.07}{\ 7/2_3^- \ (1.1002)}\connect{13.4}{1.1}{13.72}{1.07}

\end{picture}
\end{center}
\caption{Selected low-lying states of $^{135}$Sn, $^{137}$Sn, $^{139}$Sn and $^{141}$Sn. All entries are in MeV.
Boldfaced numbers represent  states
that develop or have a  large one-quasiparticle content. For energies below 1.5 MeV we list up to three states
with the same spin for $J^{\pi}=1/2^-,3/2^-,5/2^-,7/2^-,9/2^-$, since these are also discussed in connection with 
our seniority analysis. 
For higher spins we list only the Yrast states.}
\label{fig:odd_spectra}
\end{figure}
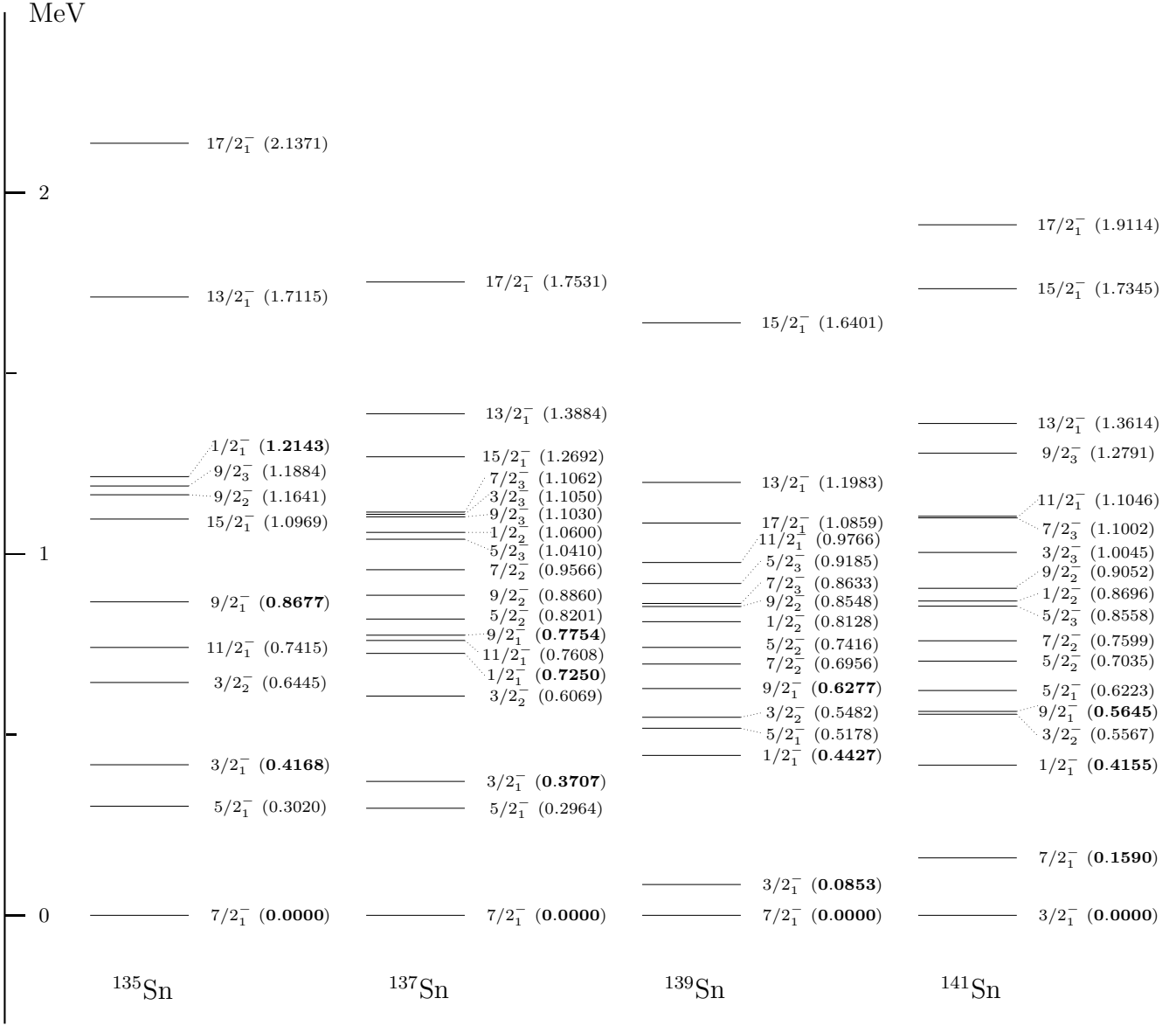

The results of the shell-model calculation for the odd isotopes are displayed in Fig.~\ref{fig:odd_spectra}.
In our pairing analysis below we show that, except for the ground states and the lowest-lying $3/2^-$ states, most of the other states
displayed in Fig.~\ref{fig:odd_spectra} deviate from a one-quasiparticle picture. We postpone therefore an analysis of the odd isotopes
to the seniority analysis below, see subsection \ref{subsec:gensensec}.
In connection with the seniority analysis, we would however like to guide the eye to the evolution of the 
$(1/2)^-$, $(3/2)^-$, $(5/2)^-$, $(7/2)^-$ and 
$(9/2)^-$ states as functions of $A$. These states can be well represented by a one-quasiparticle picture, and the energies of these
states plunge down in the spectrum with increasing $A$, a feature shown in 
Fig.~\ref{fig:odd_spectra}. States with a one-quasiparticle character are shown with bold values for the energies. Other states, such as the first $(1/2)^-$ and $(5/2)^-$, which also plunge down in the spectrum with increasing $A$,
are more fragmented  in $^{135}$Sn and $^{137}$Sn, as discussed in subsection \ref{subsec:oddseniority}. 
The $(11/2)^-$ and $(15/2)^-$ states are more complex many-body states.

For the sake of completeness, we display also selected 
calculating the $B(E2)$
transition probabilities from the first excited state to the ground state in Table \ref{tab:E2_odd_isotopes}.
We have used the effective neutron charge value of $e_{n}^{eff} = 0.64\ e$. In the calculations of $^{141}$Sn we
limited the number of particles which can occupy the $0i_{13/2}$ orbital to 9.
The shell closure in $^{140}$Sn is reflected in these transitions as well.

\begin{table}[htbp]
\begin{center}
\caption{$B(E2)$ for $^{135}$Sn, $^{137}$Sn, $^{139}$Sn and $^{141}$Sn for the lowest states. $e_{n}^{eff}=0.64e$. 
All entries are in W.u.}
\begin{tabular}{cc cc cc cc cc}
\hline
\multicolumn{2}{c}{Transition} & 
\multicolumn{2}{c}{$^{135}$Sn} & 
\multicolumn{2}{c}{$^{137}$Sn} & 
\multicolumn{2}{c}{$^{139}$Sn} &
\multicolumn{2}{c}{$^{141}$Sn} \\
\hline
&$(3/2)_1^-  \rightarrow (7/2)_1^-$   && 1.66  &&  0.27  &&  0.005 &&      \\
&$(7/2)_1^-  \rightarrow (3/2)_1^-$   &&       &&        &&        && 0.16 \\
\hline
\end{tabular}
\label{tab:E2_odd_isotopes}\end{center}
\end{table}

\subsection {Generalized seniority}\label{subsec:gensensec}

We will here interpret the shell model results in terms of the generalized seniority
model \cite{talmi1993}. The generalized seniority scheme is an extension of the seniority
scheme, that is from involving only one single orbital in $j$, the model is generalized
to involve a group of $j$-orbitals within a major shell. The generalized seniority
scheme is a simpler model than the shell model since a rather limited
number of configurations with a strictly defined structure are included. 
These states can also be given a straightforward 
physical interpretation since they represent only selected correlations.
If we by closer investigation and comparison of the shell-model wave function and the
seniority states find that the most important components are accounted for
by the seniority scheme, we can benefit from this and reduce the shell-model basis.
This would be particularly useful when we want to do calculations of systems
with a large number of valence particles.
For an even system, 
states with seniority $\nu = 0$ are
by definition states where all particles are paired. Seniority $\nu = 2$
states have one pair broken, seniority $\nu = 4$ states have two pairs broken, etc.
The admixture of seniority $\nu=4$ 
states into seniority $\nu=0$  and $\nu=2$  states
was analysed by for example \cite{Bonsignori85,Allaart88}. It was  found that
for some states  the  admixture from seniority $\nu = 4$  states
could be as large as  $20\%$ .

   Outside the closed shell, in our case $^{132}$Sn, we construct a
    generalized seniority $\nu = 0$ pair
 \begin{equation}
      S^+ = \sum_j \frac{1}{\sqrt{2j + 1}}\ C_j \sum_{m \ge 0} (-1)^{j-m} 
  a^+_{jm} a^+_{j-m}
  \end{equation}
   and identify the coefficients $C_j$ in the two-particle wave function  $S^{+}|0>$ with 
   the calculated ground state amplitudes of the complete shell-model state $^{134}$Sn.
   Similarily, the generalized seniority $\nu=2$ operators take the form
  \begin{equation}
      D^+_{\mu J} \, = \! \sum_{j \le j^\prime, m \ge 0} \frac{1}{\sqrt {(1 + 
  \delta_{j,j^\prime})}} \,\, \beta_{j,j^\prime}^{\mu J} \, \langle j m j^\prime 
  -\!\!m \vert J 0 \rangle \,\,  a^+_{jm} a^+_{j^\prime-m}
  \end{equation}
  where the coefficients $\beta_{j,j^\prime}^{\mu J}$ are obtained 
  by identifying the two-particle wave function  $D^+_{\mu J}|0>$ 
with the calculated amplitudes of the excited states of $^{134}$Sn. Note that excited
$J = 0^{+}$ states are intepreted as  generalized seniority $\nu=2$ states.

In this framework  we calculate 
the even $n$-particle model states as seniority $\nu = 0$:
$ \quad \left (S^{+}\right )^{n/2}|0>$ and  as seniority $\nu = 2$:
$ \quad D^+_{\mu JM}\left (S^{+}\right )^{(n-2)/2}|0>$ states. 
The odd $n$-particle model states are calculated as seniority $\nu = 1$
$\quad  a^+_{jm}\left (S^{+}\right )^{(n-1)/2}|0>$ states.
The squared overlaps between the constructed generalized 
  seniority states and our complete 
  shell model states, labelled $SM$ are given in the equations below.
 
For the even Sn 
  isotopes we evaluate the seniority $\nu = 0$ and $\nu = 2$
  square overlaps 
\begin{equation}
\label{eq:overlap1}
  {\cal O}_{J=0}^{\nu=0} = \left | \, \langle \mathrm{^ASn(SM)}; 0^+
  \vert \, (S^+)^{\frac{n}{2}} \, \vert \, \tilde 0 \, \rangle \, \right |^2
  \end{equation}
  and 
\begin{equation}
\label{eq:overlap2}
  {\cal O}_{J_i}^{\nu=2} = \left |\langle \mathrm{^ASn(SM)}; J_i \,\,
  \vert \, D^+_{J}(S^+)^{\frac{n - 2}{2}} \, \vert \, \tilde 0 \rangle
  \right |^2
  \end{equation}
  where the $D^+_{\mu J}$ operator corresponding to the lowest state of a given $J$ in
$^{134}$Sn has been used. For the odd Sn isotopes, the seniority square overlaps are evaluated according 
to
  \begin{equation}
\label{eq:overlap3}
  {\cal O}_{J_i}^{\nu=1} = \left | \langle \mathrm{^ASn(SM)}; J_i \,\, \vert 
  \, (S^+)^{\frac{n-1}{2}} a^+_{J_i\frac{1}{2}}\, \vert \, \tilde 0 \, 
  \rangle \right |^2 \\[.3cm]
  \end{equation}
  The vacuum state $\vert \, \tilde 0 \rangle$ is the $^{132}$Sn core, 
  $n$ is the number of valence
  neutrons and the index $i$ labels the different shell-model states with the 
  same $J$.
  These overlaps are 
  thus a measure of the strength of generalized pairing correlations of either the ground 
  state or various excited states.
  
    \subsection{Even isotopes}
  
  A typical feature of the seniority scheme is that the spacing of energy 
  levels are independent
  of the number of valence particles. Our results for the even-mass tin 
  isotopes, see Table \ref{tab:even_isotopes}, show that the spacing between the ground 
  state and the $2_1^+$
  state is fairly constant throughout the whole sequence of isotopes. This 
  suggests that the
  $2_1^+$ states may be described as one broken pair (seniority $\nu = 2$) 
  upon a ground state
  condensate of $0^+$ pairs.
  
  \begin{table}[h]
  \begin{center}
  \caption{The seniority $\nu\!=\!0\,$ squared overlap $\,{\cal 
  O}_{J=0}^{\nu=0}\,$ for the
  ground states and seniority $\nu\!=\!2\,$ squared overlaps $\, {\cal 
  O}_{J_i}^{\nu=2}\,$ for the
  lowest-lying eigenstates of the even Sn-isotopes.}
  \begin{tabular}{ccccc ccccc ccccc ccccc ccccc}
  \hline
    \multicolumn{5}{c}{$J^{\pi}_i$} & \multicolumn{5}{c}{$^{136}$Sn} & 
  \multicolumn{5}{c}{$^{138}$Sn} &
  \multicolumn{5}{c}{$^{140}$Sn} &
  \multicolumn{5}{c}{$^{142}$Sn} \\
  \hline
    &&  $0_1^+$   &&&   &&  0.967  &&&  &&  0.929  &&&  && 0.879  &&&  && 0.758 &&\\
  &&  $2_1^+$   &&&   &&  0.889  &&&  &&  0.824  &&&  && 0.743  &&&  && 0.549 &&\\
  &&  $4_1^+$   &&&   &&  0.586  &&&  &&  0.728  &&&  && 0.251  &&&  && 0.251 &&\\
  &&  $4_2^+$   &&&   &&  0.314  &&&  &&  0.000  &&&  && 0.174  &&&  && 0.046 &&\\
  &&  $6_1^+$   &&&   &&  0.896  &&&  &&  0.702  &&&  && 0.657  &&&  && 0.003 &&\\
  &&  $6_2^+$   &&&   &&  0.000  &&&  &&  0.000  &&&  && 0.014  &&&  && 0.410 &&\\
  &&  $8_1^+$   &&&   &&  0.002  &&&  &&  0.678  &&&  && 0.573  &&&  && 0.430 &&\\
  &&  $8_2^+$   &&&   &&  0.843  &&&  &&  0.001  &&&  && 0.023  &&&  && 0.016 &&\\
     \hline
  \end{tabular}
  \label{tab:overlaps_even_isotopes}
  \end{center}
  \end{table}
The results displayed in Table \ref{tab:overlaps_even_isotopes} indicate that pairing effects are 
likely to be important. The $0^+$ ground states and the first excited $2^+$ 
states have indeed large seniority $\nu = 0$ and $\nu = 2$ components, 
respectively. Not unexpectedly, however, these components are gradually
decreasing as the number of valence neutrons is increasing. This 
suggests that the pairing effects become less dominant as one approaches 
the neutron drip line. A slight exception is seen for the lowest $4^+$ 
states, for which the pairing strength is strongly fragmented in $^{136}$Sn, 
then concentrated again in $^{138}$Sn before being further fragmented in 
$^{140}$Sn and $^{142}$Sn.  
    
Our calculations also show that (not listed in the above table), in all even isotopes considered, one of the low-lying
$0^+$ excited states contains a relatively large seniority $\nu=2$ component.

Note that in evaluating the overlaps of Eqs.~(\ref{eq:overlap1}) and (\ref{eq:overlap2}) 
for Table \ref{tab:overlaps_even_isotopes} we used $S^+$ 
and $D^+$ operators determined from the two-particle wave functions of 
$^{134}$Sn and kept constant throughout the shell. This is consistent with 
both our many-body shell-model approach and the general seniority 
approach of Talmi. Thus, we have refrained from adjusting the 
coefficients of Eqs.~(\ref{eq:overlap1}) and (\ref{eq:overlap2}) as valence pairs are added, despite 
the fact that this could have given larger overlaps beyond $^{136}$Sn.
  
It is also interesting to point out the behavior of the two $8^+$ states listed in Table 
\ref{tab:even_isotopes}. 
The first $8^+$ state in $^{134}$Sn is mainly $0h_{9/2}1f_{7/2}$ while the  
second $8^+$ state in $^{134}$Sn is almost purely $(0h_{9/2})^2$ and comes at a much higher 
energy. The lowest $8^+$
changes character in $^{136}$Sn to a state dominated by the $(1f_{7/2})^4$ configuration, which in 
a seniority language corresponds to a seniority $\nu=4$ state. It is first in $^{138}$Sn that seniority $\nu=2$ has a large 
overlap with the lowest shell-model state, with the configuration $(1f_{7/2})^30h_{9/2}$ as the dominating one.

  \subsection{Odd isotopes}\label{subsec:oddseniority}

  \begin{table}[h]
  \begin{center}
  \caption{The seniority $\nu\!=\!1\,$ squared overlaps $\,{\cal 
  O}_{J_i}^{\nu=1}$ for the
  lowest-lying eigenstates of the odd Sn-isotopes. }
  \begin{tabular}{c cccc cccc cccc ccc}
  \hline
   \multicolumn{2}{c}{ $J^{\pi}_i$} & \multicolumn{3}{c}{ $^{135}$Sn} &
  \multicolumn{3}{c}{  $^{137}$Sn} & \multicolumn{3}{c}{ $^{139}$Sn} &
                                      \multicolumn{3}{c}{ $^{141}$Sn}\\
  \hline
   7/2$_1^-$ &&  & 0.966 &&  & 0.896 &&  & 0.813 &&  & 0.729  &&\\
    3/2$_1^-$ &&  & 0.005 &&  & 0.773 &&  & 0.709 &&  & 0.689 &&\\
   3/2$_2^-$ &&  & 0.936 &&  & 0.088 &&  & 0.120 &&  & 0.066 &&\\
    9/2$_1^-$ &&  & 0.003 &&  & 0.019 &&  & 0.580 &&  & 0.552 &&\\
   9/2$_2^-$ &&  & 0.176 &&  & 0.608 &&  & 0.007 &&  & 0.002 &&\\
   9/2$_3^-$ &&  & 0.560 &&  & 0.000 &&  & 0.004 &&  & 0.040  &&\\
    1/2$_1^-$ &&  & 0.631 &&  & 0.277 &&  & 0.368 &&  & 0.660 &&\\
   1/2$_2^-$ &&  & 0.223 &&  & 0.341 &&  & 0.318 &&  & 0.018 &&\\
   1/2$_3^-$ &&  & 0.133 &&  & 0.138 &&  & 0.083  &&  & 0.030   &&\\
    5/2$_1^-$ &&  & 0.006 &&  & 0.002 &&  & 0.008 &&  & 0.168 &&\\
   5/2$_2^-$ &&  & 0.272 &&  & 0.030 &&  & 0.369 &&  & 0.357 &&\\
   5/2$_3^-$ &&  & 0.041 &&  & 0.308 &&  & 0.012 &&  & 0.021 &&\\
   5/2$_4^-$ &&  & 0.172 &&  & 0.004 &&  & 0.005 &&  & 0.003&&\\
   \hline
  
  \end{tabular}
  \label{tab:odd_seniority}
  \end{center}
  \end{table}
  
For the odd-mass Sn isotopes the seniority $\nu = 1$ square overlaps of Eq.~(\ref{eq:overlap3}) 
are displayed in Table \ref{tab:odd_seniority}. These quantities allow us to see how the 
strengths of the various single-quasiparticle levels of  $^{133}$Sn evolve as 
valence neutrons are added. We observe that the single-particle strength 
is largely preserved for the low-lying $9/2^-$, $7/2^-$ and $3/2^-$ states, although 
being slowly attenuated with increasing number of valence nucleons. For 
higher lying states such as the $5/2^-$ states, the strength is more strongly fragmented due to 
mixing with neighboring states. Comparing Table \ref{tab:odd_seniority} with Fig.~\ref{fig:odd_spectra}
we furthermore see that the higher single-quasiparticle 
levels are shifted downwards as neutron pairs are being added in the 
$f_{7/2}$ level, signaling again the importance of pair correlations in these 
nuclei.
Thus, the $3/2^-$ quasiparticle state, which is at 0.64 MeV in $^{135}$Sn and at 
0.37 MeV in $^{137}$Sn becomes nearly degenerate with the $7/2^-$ ground state 
in $^{139}$Sn and takes over as ground state in $^{141}$Sn.  
For the $9/2^-$ states, we notice that for $^{135}$Sn, it is the second excited state which has the largest
seniority $\nu=1$ overlap. For  $1/2^-$ there is  a lowest state with a large one-quasiparticle
overlap except for  $^{137}$Sn.

The large one-quasiparticle overlaps are also reflected in the fact that the energies of the corresponding
states come down in energy as $A$ increases, a feature displayed in Fig.~\ref{fig:odd_spectra} with boldfaced numbers for the states
with a large one-quasiparticle content.

\section{Conclusions}\label{conclusions}

We have performed extensive shell model studies of the heavy tin isotopes with up to ten valence neutrons outside
the closed $^{132}$Sn core. A realistic microscopic effective interaction has been derived from a modern meson exchange
NN potential using many-body perturbation theory. A generalized seniority analysis was applied in order to get insights on
the pairing structure of the shell model states. The results of our calculations are in good agreement with the few
available experimental data. The other results can serve as a guideline for future experiments.
Furthermore, a seniority analysis shows that pairing correlations are strong even for 
systems with ten valence particles, although the strength of the pairing correlations gets reduced as we increase the number
of valence neutrons. Our results, especially for  even nuclei, demonstrate that a seniority approach is
a good starting point for shell-model studies of heavy Sn isotopes, as observed for the lighter tin isotopes from $^{102}$Sn to 
$^{130}$Sn as well.


\begin{thebibliography}{200}
\bibitem{bhattacharyya2001}P.~Bhattacharyya {\em et al}, Phys.~Rev.~Lett.~{\bf 87}, 062502 (2001).
\bibitem{radford2002} D.~C.~Radford {\em et al}, Phys.~Rev.~Lett.~{\bf 88}, 222501 (2002). 
\bibitem{hribf} HRIBF newsletter, July 2003,
http://www.phy.ornl.gov/hribf/usersgroup/news/jul-03/jul-03.html\#B
\bibitem{dillman2003}  I. Dillman {\em et al}, Phys.~Rev.~Lett.~{\bf 91}, 162503 (2003). 
\bibitem{jacob2002} G.~Jacob {\em et al}, Phys.~Rev.~C {\bf 65}, 024316 (2002).
\bibitem{terasaki2002} J.~Terasaki, J.~Engel, W.~Nazarewicz, and M.~Stoitsov, Phys.~Rev.~C {\bf 66}, 054313 (2002).
\bibitem{stone2005} N.~J.~Stone {\em et al}, Phys.~Rev.~Lett.~{\bf 94}, 192501 (2005).
\bibitem{shergur2002} J.\ Shergur {\em et al}, Phys.~Rev.~C {\bf 65}, 034313 (2002). 
\bibitem{shergur2005a} J.~Shergur {\em et al}, Phys.~Rev.~C {\bf 72}, 024305 (2005). 
\bibitem{shergur2005b} J.~Shergur {\em et al}, Phys.~Rev.~C {\bf 71}, 064321 (2005).
\bibitem{is378} Proposal CERN-INTC-2003-04.
\bibitem{hoff96}P.~Hoff {\em et al}, Phys.\ Rev.\ Lett.~{\bf 77}, 1020 (1996). 
\bibitem{gigi2002} L.~Coraggio, A.~Covello, A.~Gargano, and N.~Itaco, Phys.~Rev.~C {\bf 65}, 051306 (2002).
\bibitem{hoffmann} F.~Hofmann and H.~Lenske, Phys.~Rev.~C {\bf 57}, 2281 (1998).
\bibitem{doba1996} J.~Dobaczewski, W.~Nazarewicz, T.~R.~Werner, J.~F.~Berger, C.~R.~Chinn, and J.~Decharge,  
Phys.~Rev.~C {\bf 53}, 2809 (1996).    
\bibitem{talmi1993} I.~Talmi, {\em Simple Models of Complex Nuclei},
Contemporary Concepts in Physics (Harwood Academic Publishers, Chur, Switzerland, 1993) Vol. 7.
\bibitem{Bonsignori85}G.\ Bonsignori, M.\ Savoia, K.\ Allaart,
A.\ van Egmond and
G.\ Te Velde, Nucl.\ Phys.\ {\bf A432}, 389 (1985).
\bibitem{Allaart88}K.\ Allaart, E.\ Boeker, G.\ Bonsignori,
M.\ Savoia and Y.\ K.\ Gambhir,
Phys.\ Rep.\ {\bf 169}, 211 (1988).
\bibitem{hko95}  M.~Hjorth-Jensen, T.~T.~S.~Kuo and E.~Osnes, Phys.~Rep.~{\bf 261}, 125 (1995).
\bibitem{cdbonn} R.~Machleidt, F.~Sammarruca and Y.~Song, Phys.~Rev.~C {\bf 53}, R1483 (1996).
\bibitem{mac89}  R.~Machleidt, Adv.~Nucl.~Phys.~{\bf 19}, 189 (1989). 
\bibitem{brown2005} B.~A.~Brown, N.~J.~Stone, J.~R.~Stone, I.~S.~Towner, and M.~Hjorth-Jensen,
Phys.~Rev.~C {\bf 71}, 044317 (2005).
\bibitem{stone1998} G.~N.~White {\em et al}, Nucl.\ Phys.~{\bf A644}, 277 (1998). 
\bibitem{isotones1997} A.~Holt, T. Engeland, E. Osnes, M.~Hjorth-Jensen, and J.~Suhonen, Nucl.\ Phys.~{\bf A618}, 107 (1997). 
\bibitem{whit77} R.~R.~Whitehead, A.~Watt, B.~J.~Cole  and I.~Morrison, Adv.~Nucl.~Phys.~{\bf 9}, 123 (1977).
\bibitem{torgeir1991} T.~Engeland, the Oslo shell model code, 1991-2006, unpublished. 
\bibitem{urban99}W.~Urban {\em et al},  Eur.~Phys.~J. A {\bf 5}, 239 (1999). 
\bibitem{mezilev} K.~A.~Mezilev, Yu.~N.~Novikov, A.~V.~Popov, B.~Fogelberg, and L.~Spanier, Phys.~Scripta T {\bf 56}, 272 (1995).
\bibitem{jpg22} M.~Hjorth-Jensen, H.~M\"uther, E.~Osnes, and A.~Polls, J.~Phys.~{\bf G22}, 321 (1996).
\bibitem{korgul2000} A.~Korgul {\em et al}, Eur. Phys.~J.~{\bf A7}, 167 (2000).
\bibitem{eivind1965} E.~Osnes, PhD thesis, University of Oslo, (1966), unpublished.
\bibitem{eivind1966} T.~Engeland and E.~Osnes, Phys.~Lett.~{\bf 20}, 424 (1966). 
\bibitem{dhj2003} D.~J.~Dean and M.~Hjorth-Jensen, Rev.~Mod.~Phys.~{\bf 75}, 607 (2003).
\bibitem{bm69} A.~Bohr and B.~R.~Mottelson, {\em Nuclear Structure}, (World Scientific, Singapore, 1998), Vol.~{\bf 1}. 
\bibitem{zhang1997} C.T.\  Zhang {\em et al}, Z.~Phys.~{\bf A358}, 7 (1997).
\end{thebibliography}
\end{document}